\documentclass[twocolumn,showpacs,preprintnumbers,amsmath,amssymb]{revtex4}

\usepackage{graphicx}
\usepackage{dcolumn}
\usepackage{bm}

\begin{document}


\title{First-order phase transition in the tethered surface model on a sphere}

\author{Hiroshi Koibuchi}
 \email{koibuchi@mech.ibaraki-ct.ac.jp}

\author{Toshiya Kuwahata}%

\affiliation{%
Department of Mechanical and Systems Engineering, Ibaraki College of Technology, Nakane 866 Hitachinaka, Ibaraki 312-8508, Japan}%


\begin{abstract}
We show that the tethered surface model of Helfrich and Polyakov-Kleinert undergoes a first-order phase transition separating the smooth phase from the crumpled one. The model is investigated by the canonical Monte Carlo simulations on spherical and fixed connectivity surfaces of size up to $N\!=\!15212$. The first-order transition is observed when $N\!>\!7000$, which is larger than those in previous numerical studies, and a continuous transition can also be observed on small-sized surfaces. 
Our results are, therefore, consistent with those obtained in previous studies on the phase structure of the model. 
\end{abstract}

\pacs{64.60.-i, 68.60.-p, 87.16.Dg}
\maketitle

\section{Introduction}\label{intro}
Considerable progress has been made in understanding the phase structure of the model of Helfrich and Polyakov-Kleinert \cite{HELFRICH-1973,POLYAKOV-NPB1986,NELSON-SMMS2004,David-TDQGRS-1989,Wiese-PTCP2000,Bowick-PREP2001,WHEATER-JP1994,Peliti-Leibler-PRL1985,David-EPL1986,DavidGuitter-EPL1988,BKS-PLA2000,BK-PRB2001}; we will abbreviate this to the HPK model. The HPK model describes not only two-dimensional surfaces swept out by a rigid string \cite{WHEATER-JP1994} but also biological membranes such as human red-blood cells and artificial vesicles \cite{NELSON-SMMS2004,David-TDQGRS-1989,Wiese-PTCP2000,Bowick-PREP2001}. 

It is now widely accepted that the tethered model of HPK undergoes a continuous phase transition. 
The Hamiltonian of the HPK model is given by a linear combination of the Gaussian term $S_1$ and the bending energy term $S_2$: $S\!=\! S_1 \!+\!bS_2$, where $b$ is the bending rigidity. $S_2$ is ordinarily defined by using the unit normal vectors for the triangles. The large-$D$ expansion \cite{DavidGuitter-EPL1988} predicts that the HPK model undergoes a finite-$b$ continuous transition between the smooth phase in the limit $b\!\to\!\infty$ and the crumpled phase in the limit $b\!\to\!0$. Numerical studies performed so far have also focused on the phase transition in the tethered surface models of HPK \cite{KANTOR-NELSON-PRA1987, KANTOR-SMMS2004, WHEATER-NPB1996, BCFTA-JP96-NPB9697, KY-IJMPC2000-2, KOIB-PLA2003-2} and indicate that the model exhibits the continuous transition. 

On the contrary, we can also think that the model has a discontinuous transition. It was predicted by mean field analysis that the model undergoes a first-order phase transition \cite{PKN-PRL1988}. In recent numerical simulations on fixed connectivity surfaces \cite{KD-PRE2002,Koibuchi-PRE-2004-1}, it was also suggested that the phase transition is of the first order. The Hamiltonian of the model in \cite{KD-PRE2002} includes the Lennard-Jones potential serving as the Gaussian energy for the HPK model. The bending energy in \cite{Koibuchi-PRE-2004-1} is very similar to the one for the HPK model. These numerical results, therefore, strongly suggest that the HPK model undergoes a discontinuous transition.

However, little attention has been given to whether a discontinuous transition is observed in the tethered surface model of HPK, whose Hamiltonian includes the bending energy of the form $1\!-\!{\bf n}_i\cdot {\bf n}_j$,  where ${\bf n}_i$ is the unit normal vector for the triangle $i$. We will call this form of energy as the ordinary bending energy from now on. Therefore, in order to confirm that the phase transition of the HPK model is of the first order, we need to study further the tethered surface model defined by the ordinary bending energy. 

In this paper, we numerically study the tethered model on a sphere, whose Hamiltonian is given by $S\!=\!S_1\!+\!bS_2$, where $S_1$ is the Gaussian energy and $S_2$ is the ordinary bending energy described above. This Hamiltonian has been widely accepted and investigated as a discrete model of HPK. Although a fluid surface model  \cite{CATTERALL-NPBSUP1991, AMBJORN-NPB1993, ABGFHHM-PLB1993, BCHHM-NPB9393,KY-IJMPC2000-1, KOIB-PLA200234} defined on dynamically triangulated surfaces is very interesting and should be investigated further on larger surfaces, we will concentrate on the fixed connectivity tethered model in this paper. 

We will show the first numerical evidence that the ordinary tethered model undergoes a discontinuous transition on a sphere. The gap of the bending energy is clearly seen on surfaces of $N\!\geq\! 7000$, and cannot be seen on the smaller surfaces. It must be emphasized that the results are not contradictory to previous ones, as the continuous transition can also be observed in our simulations on smaller surfaces. 

\section{The Model and Monte Carlo technique}
The partition function of the model is defined by
\begin{eqnarray}
\label{part-func}
&&Z = \int \prod_{i=1}^N d X_i \exp\left[ -S(X) \right] , \nonumber \\
&&S(X)=S_1 + b S_2 
\end{eqnarray}
where $b$ is the bending rigidity, $N$ the total number of vertices. The center of the surface is fixed to remove the translational zero mode. The self-avoiding property of surfaces is not assumed in the integrations $d X_i$ in ${\bf R}^3$. The symbols $S_1$, $S_2$  in Eq. (\ref{part-func}) denote the Gaussian energy and the bending energy, which were already introduced in the Introduction and are defined by 
\begin{equation}
\label{S1S2}
S_1=\sum_{(i,j)} \left(X_i-X_j\right)^2,\; S_2=\sum_{i,j}\left(1-{\bf n}_i\cdot{\bf n}_j \right),
\end{equation}
where $\sum_{(i,j)}$ denotes the sum over bonds $(i,j)$, and $\sum_{i,j}$ the sum over triangles $i,j$ sharing a common bond. The symbol ${\bf n}_i$ in Eq. (\ref{S1S2}) denotes the unit normal vector of the triangle $i$, as was introduced in the Introduction. 

The canonical Monte Carlo (MC) technique is used to update the variables $X$. The new position $X^\prime_i$ of the vertex $i$ is given by $X^\prime_i\!=\!X_i\!+\!{\it \Delta} X$, where ${\it \Delta X}$ is chosen randomly in a small sphere. The radius of the small sphere is chosen at the start of the MC simulations to maintain about a $50\%$ acceptance rate. The new position $X^\prime_i$ is accepted with the probability ${\rm Min}[1,\exp\left(-{\it \Delta}S\right) ]$, where ${\it \Delta}S$ is given by ${\it \Delta}S\!=\!S({\rm new})\!-\!S({\rm old})$.

The minimum bond length is not assumed. On the contrary, the minimum area of triangle is assumed to be $10^{-6}\times A_0$, where $A_0$ is the mean area of triangles computed at every 250 Monte Carlo sweeps (MCS) and is almost constant throughout the MC simulations. The area of almost all triangles generated in the MC simulations is larger than the lower bound $10^{-6}\times A_0$. 

We use a random number called Mersenne Twister \cite{Matsumoto-Nishimura-1998} in the MC simulations. Two sequences of random number are used; one for a 3-dimensional move of vertices $X$ and the other for the Metropolis accept/reject for the update of $X$.

The surfaces, on which the Hamiltonian in Eq. (\ref{S1S2}) is defined, are obtained by dividing the icosahedron, and hence, are uniform in the co-ordination number. By dividing every edge of the icosahedron into $L$ pieces of the same length, we have a triangulated surface of size $N\!=\!10L^2\!+\!2$. These surfaces are thus characterized by $N_5\!=\!12$ and $N_6\!=\!N\!-\!12$, where $N_q$ is the total number of vertices with co-ordination number $q$. Hence we have surfaces in which 12 vertices are of $q_i\!=\!5$, and all other vertices $q_i\!=\!6$. Hence, the surfaces are made uniquely in contrast to the Voronoi lattices constructed by using random numbers.

We comment on the unit of physical quantities. The scale of length in the model can be chosen arbitrarily because of the scale invariant property of the partition function in Eq. (\ref{part-func}). Then, by letting $a$ be a length unit (the mean bond length for example) in the model, we can express all quantities with unit of length by $a$, which is assumed to be $a\!=\!1$ in the model. Hence, the unit of $S_1$ is $a^2$. Let $\lambda$ be the surface tension coefficient, which is assumed to be $\lambda\!=\!1$,  $S$ in Eq. (\ref{part-func}) can also be written as $S\!=\!\lambda S_1\!+\! b S_2$. Thus, the unit of $\lambda$ can be written as $kT/a^2$, where $k$ is the Boltzmann constant and $T$ is the temperature. The unit of $b$ is then expressed by $kT$.

\section{Results}
\begin{figure}[hbt]
\centering
\includegraphics[width=8.5cm]{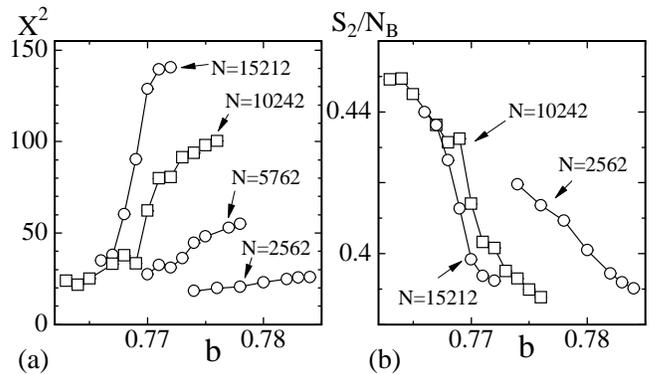}
 \caption{(a) The mean square size $X^2$ against the bending rigidity $b$, and (b) the bending energy $S_2/N_B$ against $b$, where $N_B$ is the total number of bonds. Both $X^2$ and $S_2/N_B$ discontinuously change at some intermediate $b$. This discontinuity represents some discontinuous phase transition. The unit of $X^2$ is $a^2$, where $a$ is the length unit in the model, which is chosen to be $a\!=\!1$.} 
\label{fig-1}
\end{figure}
We firstly show in Fig. \ref{fig-1}(a) the mean square size $X^2$ defined by
\begin{equation}
\label{X2}
X^2={1\over N} \sum_i \left( X_i -\bar X \right)^2,\qquad \bar X ={1\over N} \sum_i X_i,
\end{equation}
where $\sum_i$ denotes the sum over vertices $i$. Discontinuous changes of $X^2$ are visible at intermediate bending rigidity $b$ in Fig. \ref{fig-1}(a), and suggest that there is a discontinuous transition between the smooth and the crumpled phases.  

Figure  \ref{fig-1}(b)  shows the bending energy $S_2/N_B$ against $b$, where $N_B$ is the total number of bonds. We find a discontinuity in $S_2/N_B$ at $b$ where $X^2$ discontinuously changes. This can be viewed as a signal of a discontinuous transition.  

Total number of MCS at the transition point was $5\!\times\!10^8$,  $8\!\times\!10^8$,  and $8\!\times\!10^8$  for surfaces of $N\!=\!7292$, $N\!=\!10242$, and $N\!=\!15212$, respectively. A relatively small number of MCS were done at $b$ far distant from the transition point on each surface.

\begin{figure}[hbt]
\centering
\includegraphics[width=8.5cm]{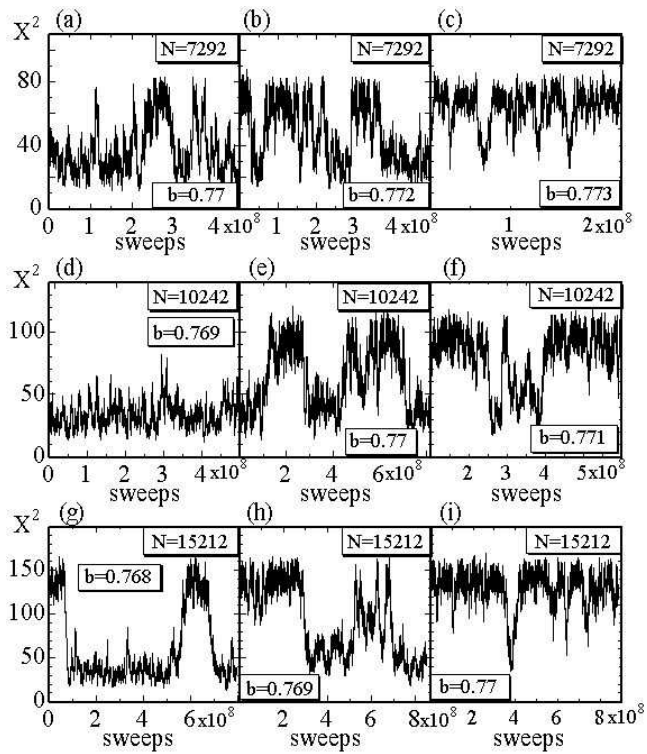}
 \caption{The variation of $X^2$ against MCS obtained on the $N\!=\!7292$ surface at (a) $b\!=\!0.77$,  (b) $b\!=\!0.772$, and (c)  $b\!=\!0.773$,  those obtained on the $N\!=\!10242$ surface at (d) $b\!=\!0.769$,  (e) $b\!=\!0.77$, and (f)  $b\!=\!0.771$, and those obtained on the $N\!=\!15212$ surface at (g) $b\!=\!0.768$,  (h) $b\!=\!0.769$, and (i)  $b\!=\!0.77$.      }
\label{fig-2}
\end{figure}
Figures \ref{fig-2}(a), (b), and (c) represent the variation of $X^2$ against MCS, which were obtained at $b\!=\!0.77$, $b\!=\!0.772$, and $b\!=\!0.773$ on the surface of size $N\!=\!7292$. We can find from Fig. \ref{fig-2}(b) that there are at least two states which differ in size; one of them is characterized by $X^2\!\simeq\!30$ and the other by $X^2\!\simeq\!70$, which correspond to a crumpled and a smooth state respectively at the transition point of the $N\!=\!7292$ surface. The size of the surface appears to be stable, and this stability of size reflects a phase transition. Thus, we understand that the surface remains in the crumpled (smooth) phase at $b\!=\!0.77$ ($b\!=\!0.773$), and that the transition point is close to $b\!=\!0.772$ on the $N\!=\!7292$ surface. Figures  \ref{fig-2}(d), (e) and (f) are those obtained at points $b\!=\!0.769$, $b\!=\!0.77$, and $b\!=\!0.771$ respectively on the surface of $N\!=\!10242$. We can see that the transition point for the $N\!=\!10242$ surface is close to $b\!=\!0.77$, where two states which differ in size can also be seen. Figures \ref{fig-2}(g), (h), and (i) are those for $N\!=\!15212$ at points $b\!=\!0.768$, $b\!=\!0.769$, and $b\!=\!0.77$, respectively. We find from the figures that the transition point for the $N\!=\!15212$ surface is close to $b\!=\!0.768$ or $b\!=\!0.769$. 

 The transition point moves left in the $b$-axis with the increasing $N$ as can be found in Figs. \ref{fig-1}(a) and (b). The reason for this is because of the size effect. We also note that two distinct peaks can be observed in the distribution (or the histogram) $h(X^2)$ of $X^2$, which are not depicted here. On the smaller surfaces of $N\!=\!5672$,  $N\!=\!4842$,  and $N\!=\!3242$, two distinct peaks can also be observed in $h(X^2)$ at each transition point. 

\begin{figure}[hbt]
\centering
\includegraphics[width=8.5cm]{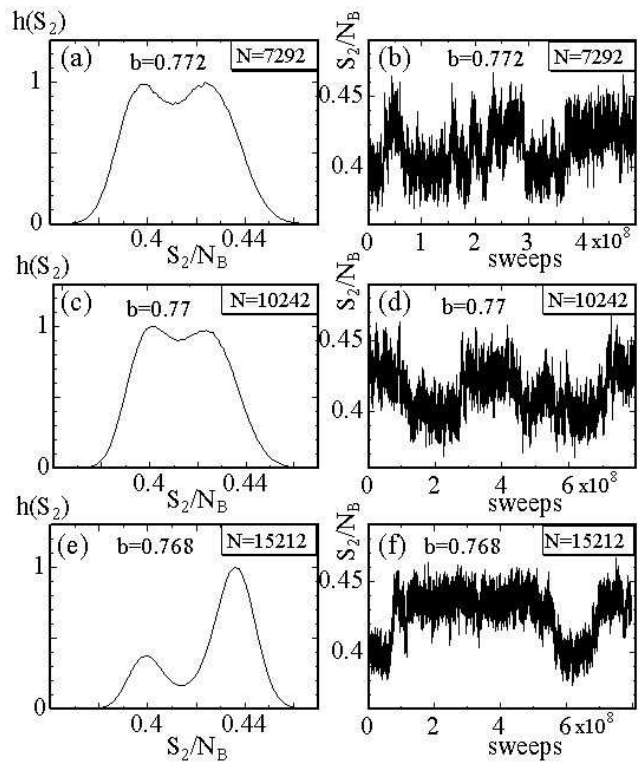}
 \caption{(a) The normalized histogram $h(S_2)$ of $S_2/N_B$, and (b) the variation of $S_2/N_B$ against MCS, on the $N\!=\!7292$ surface at $b\!=\!0.772$,  (c) $h(S_2)$ and (d) the variation of $S_2/N_B$ on the $N\!=\!10242$ surface at $b\!=\!0.77$, (e) $h(S_2)$ and (f) the variation of $S_2/N_B$ on the $N\!=\!15212$ surface at $b\!=\!0.769$. Two distinct peaks on each $h(S_2)$ indicate that the model undergoes a discontinuous transition.}
\label{fig-3}
\end{figure}
Figure \ref{fig-3}(a) is a normalized histogram $h(S_2)$ of $S_2/N_B$ obtained at the transition point $b\!=\!0.772$ on the surface of size $N\!=\!7292$. We observed two clear peaks in the $h(S_2)$ shown in Fig. \ref{fig-3}(a). Note that two distinct states, which differ in size, were observed at the point $b\!=\!0.772$ as shown in Fig. \ref{fig-2}(b). The variation of $S_2$ against MCS is plotted in Fig. \ref{fig-3}(b). Figures  \ref{fig-3}(c) and \ref{fig-3}(d) show $h(S_2)$ obtained at $b\!=\!0.77$ on the $N\!=\!10242$ surface, and  Figs. \ref{fig-2}(e) and \ref{fig-2}(f) show those of $N\!=\!15212$ at $b\!=\!0.768$. We found two distinct peaks in $h(S_2)$ shown in Figs. \ref{fig-3}(c) and \ref{fig-3}(e), just like for $N\!=\!7292$ in Fig. \ref{fig-3}(a). We found two peaks also in $h(S_2)$ of $N\!=\!15212$ at $b\!=\!0.769$, which was not presented in a figure.  The two peaks in $h(S_2)$ shown in Figs. \ref{fig-3}(a), \ref{fig-3}(c) and \ref{fig-3}(e) indicate that the model undergoes a first-order phase transition. Moreover, we find from these figures that the discontinuous nature of the transition is visible only on the surfaces of size $N\geq\!7292$. We obtained $h(S_2)$ on $N\!\leq\!5762$ surfaces and confirmed that there are no clear two-peaks in $h(S_2)$ in contrast to those shown in Figs.  \ref{fig-3}(a), \ref{fig-3}(c),  and \ref{fig-3}(e). 

In order to obtain the Hausdorff dimensions $H_{\rm smo}$ in the smooth phase close to the transition point and $H_{\rm cru}$ in the crumpled phase close to the transition point, the mean value of $X^2$ is obtained by averaging $X^2$ over a small region at each peak of $h(X^2)$:  $28\!\leq\! X^2\!\leq\!80$ and $118\!\leq\! X^2\!\leq 165$ at $b\!=\!0.769$ on the surface of $N\!=\!15212$, $20\!\leq\! X^2\!\leq\!55$ and $78\!\leq\! X^2\!\leq\!110$ at $b\!=\!0.77$ on $N\!=\!10242$, $15\!\leq\! X^2\!\leq\!45$ and $53\!\leq\! X^2\!\leq\!82$ at $b\!=\!0.772$ on $N\!=\!7292$,  $12\!\leq\! X^2\!\leq\!38$ and $40\!\leq\! X^2\!\leq\!68$  at $b\!=\!0.773$ on $N\!=\!5762$, and  $12\!\leq\! X^2\!\leq\!30$ and $35\!\leq\! X^2\!\leq\!55$  at $b\!=\!0.772$ on $N\!=\!4842$.

\begin{figure}[hbt]
\centering
\includegraphics[width=8.5cm]{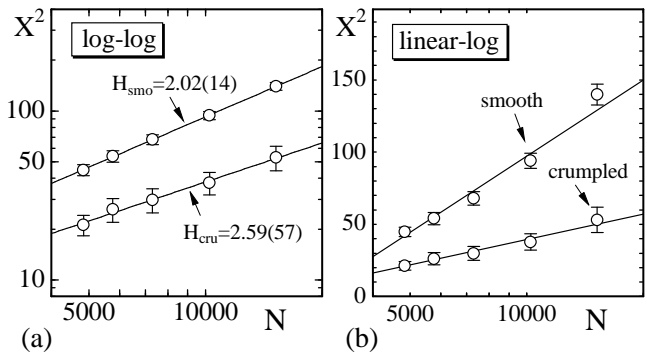}
 \caption{(a) Log-log plots of $X^2$ against $N$ obtained in the smooth phase and in the crumpled phase at the transition point of surfaces $N\!\geq\! 4842$. (b) Linear-log plots of $X^2$ against $N$. The error bars on the data are the standard deviations. }
\label{fig-4}
\end{figure}
Figure \ref{fig-4}(a) shows log-log plots of $X^2$ against $N$, which are obtained by averaging $X^2$ in the regions listed as above. Error bars on the data are the standard deviations. The straight lines are drawn in the figures by fitting the data $X^2$ to
\begin{equation}
\label{Hausdorff-scale}
X^2 \sim N^{2/H}, 
\end{equation}
and as a consequence we have 
\begin{eqnarray}
\label{H}
&&H_{\rm smo} = 2.02\pm 0.14, \nonumber \\
&&H_{\rm cru} = 2.59\pm 0.57.
\end{eqnarray}
It is remarkable that $H_{\rm cru}$ is less than the physical bound $H\!=\!3$, although the error is relatively large. We note also that $H_{\rm cru}$ in Eq. (\ref{H}) is comparable to the theoretical prediction $H=2.39(23)$ within the error,  which corresponds to the scaling exponent $\nu\!=\!0.84\!\pm\!0.04$ \cite{David-Wiese-PRL96} where $\nu\!=\!2/H$. On the contrary, $H_{\rm smo}$ is almost identical to the topological dimension 2. This indicates that the surface can be viewed as a smooth surface in the smooth phase. 

The large error in $H_{\rm cru}$ indicates that $H_{\rm cru}$ is not well defined. Hence, it is possible that $X^2$ is logarithmically divergent \cite{Gross-PLB1984,Duplantier-PLB1984,JK-PLB1984}. To check the logarithmic divergence of $X^2$ in the crumpled phase and in the smooth phase, we plot $X^2$ against $\log N$ in Fig. \ref{fig-4}(b). We immediately find from the figure that $X^2$ in the smooth phase does not scale according to $X^2\!=\!c_0\!+\!c_1 \log N$, which is expected just in the limit $b\!\to\! 0$. It is also found that $X^2$ does not scale according to the logarithmic divergence in the crumpled phase. In fact, a dimensionless quantity the residual sum of squares RSS, defined by ${\rm RSS}\!=\!\sum [({\rm data}-{\rm fitting \; formula})/{\rm error}]^2$, for $X^2$ in the crumpled phase is ${\rm RSS}\!=\!0.403$ which is obtained by the linear-log fit in Fig. \ref{fig-4}(b), and it is larger than ${\rm RSS}\!=\!0.165$ which is obtained by the log-log fit in Fig. \ref{fig-4}(a). Thus, the log-log fit is better than the linear-log fit for $X^2$ in the crumpled phase at the transition point. This allows us to conclude that $H_{\rm cru}$ in Eq. (\ref{H}) is a well-defined value. We must note, however, that only large scale simulations can clarify whether $H_{\rm cru}$ has a well-defined value and is less than the physical bound.

\begin{figure}[hbt]
\centering
\includegraphics[width=8.5cm]{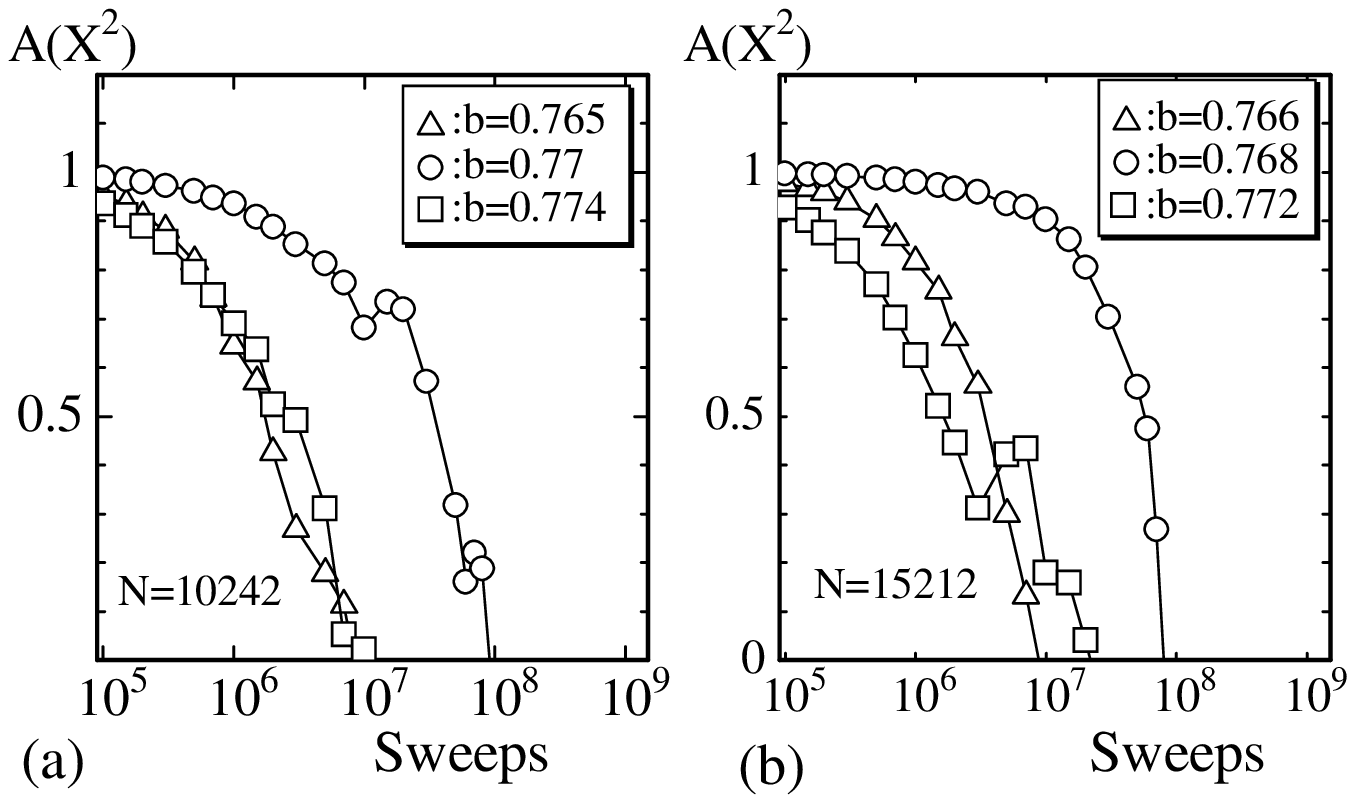}
 \caption{$A(X^2)$ obtained at the smooth phase, the crumpled phase, and the transition point on surfaces of (a)$N\!=\!10242$ and (b)$N\!=\!15212$.}
\label{fig-5}
\end{figure}

In order to see the critical slowing down typical to phase transitions, we plot in Figs. \ref{fig-5}(a) and \ref{fig-5}(b) the autocorrelation coefficient $A(X^2)$ of $X^2$ defined by 
\begin{eqnarray}
A(X^2)= \frac{\sum_i X^2(\tau_{i}) X^2(\tau_{i+1})} 
   {  \left[\sum_i  X^2(\tau_i)\right]^2 },\\ \nonumber
 \tau_{i+1} = \tau_i + n \times 500, \quad n=1,2,\cdots.  
\end{eqnarray}
The horizontal axes in the figure represent $500\!\times\! n\;(n\!=\!1,2,\cdots)$-MCS, which are a sampling-sweep between the samples $X^2(\tau_i)$ and $X^2(\tau_{i+1})$. The critical slowing down is clearly seen in the figure. The reason for this is because the volume of phase space ($\subseteq {\bf R}^3$) of $X$ at the transition point becomes larger than those at the crumpled phase and at the smooth phase. We find also an expected behavior for $A(X^2)$ such that $A(X^2)$ depends on $N$ and $A(X^2)$ becomes larger with increasing $N$, which is not plotted in the figure. 
 
\begin{figure}[hbt]
\centering
\includegraphics[width=8.5cm]{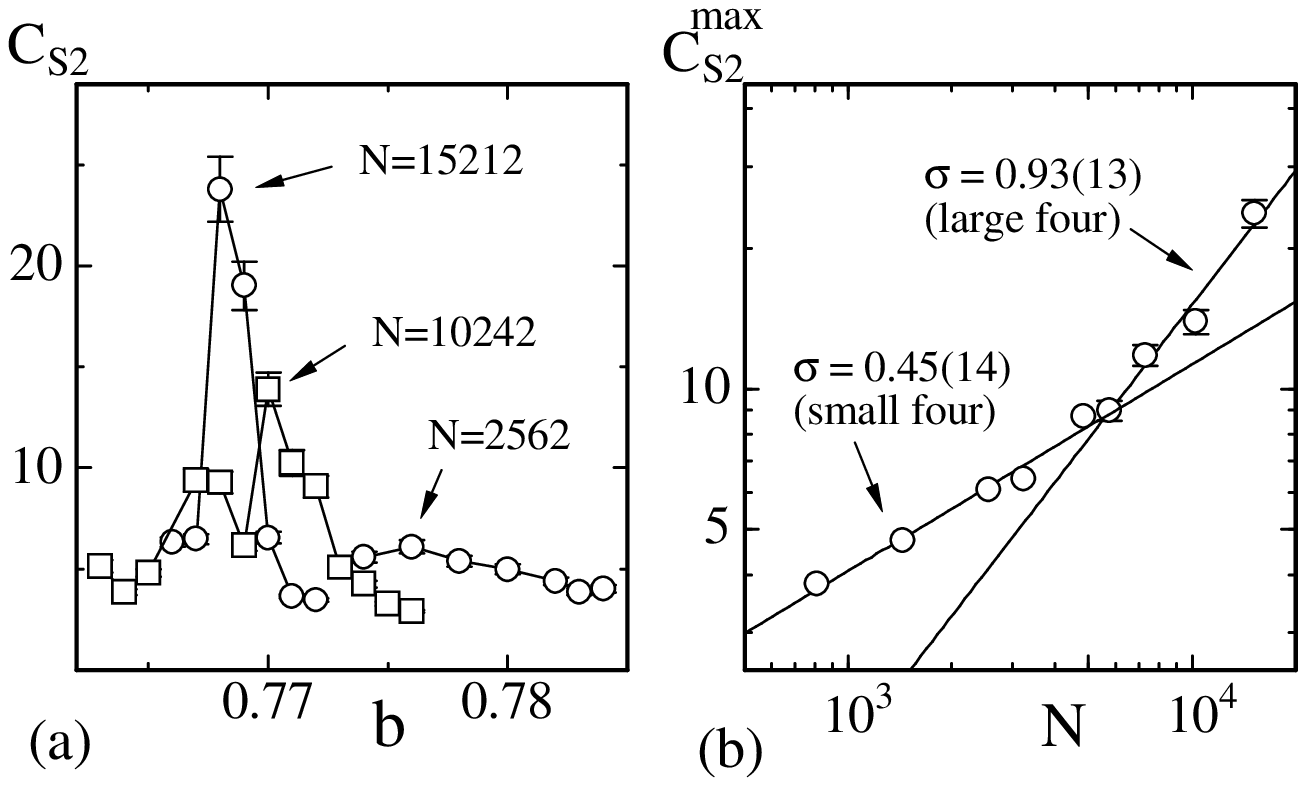}
 \caption{(a) The specific heat $C_{S_2}$ against the bending rigidity $b$, and (b) the peak values $C_{S_2}^{\rm max}$  against the number of vertices $N$ in a log-log scale. The error bars on the data are the statistical errors. The straight lines in (b) are drawn by fitting the largest four and the smallest four $C_{S_2}^{\rm max}$  to Eq. (\ref{sigma-fitting}). The units of $C_{S_2}$ and $b$ are $(kT)^2$ and $kT$ respectively.}
\label{fig-6}
\end{figure}

The specific heat $C_{S_2}$ defined by
\begin{equation}
\label{Spec-Heat}
C_{S_2} = {b^2\over N} \langle \; \left( S_2 - \langle S_2 \rangle\right)^2 \; \rangle
\end{equation}
can reflect the phase transition. Figure \ref{fig-6}(a) shows $C_{S_2}$ which were obtained on surfaces of size $N\!=\!15212$, $N\!=\!10242$, and $N\!=\!2562$ respectively. Sharp peaks of $C_{S_2}$ in Fig. \ref{fig-6}(a) indicate  a discontinuous transition. The peaks are located at $b\!=\!0.768$, $b\!=\!0.77$, $b\!=\!0.776$ on surfaces of size $N\!=\!15212$, $N\!=\!10242$, and $N\!=\!2562$ respectively. The curve of $C_{S_2}$ of $N\!=\!2562$ is relatively smooth, however, it is not so smooth for $N\!=\!15212$ or $N\!=\!10242$. These irregular behaviors for $C_{S_2}$ can be seen more or less equally for $N\!=\!7292$, $N\!=\!5762$ or $N\!=\!4842$. On these surfaces, we can see two peaks in the histograms $h(X^2)$. Thus, we understand that it is very hard to obtain $C_{S_2}$ smoothly from such surfaces whose size changes discontinuously.       

Figure \ref{fig-6}(b) is  a log-log plot of the peak value $C_{S_2}^{\rm max}$ against $N$ including the results obtained on surfaces of $N\!=\!3242$, $N\!=\!2562$, $N\!=\!1442$, and $N\!=\!812$. The straight lines were drawn by fitting the largest four $C_{S_2}^{\rm max}$ and the smallest four $C_{S_2}^{\rm max}$ to
\begin{equation}
\label{sigma-fitting}
C_{S_2} \sim N^\sigma,
\end{equation}
where $\sigma$ is a critical exponent of the transition. Thus, we have
\begin{eqnarray}
\label{sigma-result}
\sigma_1=0.93\pm 0.13,\quad (N\geq 5762),\nonumber \\
\sigma_2=0.45\pm 0.14,\quad (N\leq 4842).
\end{eqnarray}
$\sigma_1\!=\!0.93(13)$ which indicates that the phase transition is of the first order. On the contrary, $\sigma_2\!=\!0.45(14)$ in Eq. (\ref{sigma-result}) implies that the model appears to undergo a continuous transition at $N \leq 4842$. Two different behaviors of $C_{S_2}^{\rm max}$ against $N$ shown in Fig. \ref{fig-6}(b) are consistent with the fact that two distinct peaks in $h(S_2)$ can be observed only on larger ($N\!\geq\! 7292$) surfaces. 

\begin{figure}[htb]
\centering
\includegraphics[width=8.5cm]{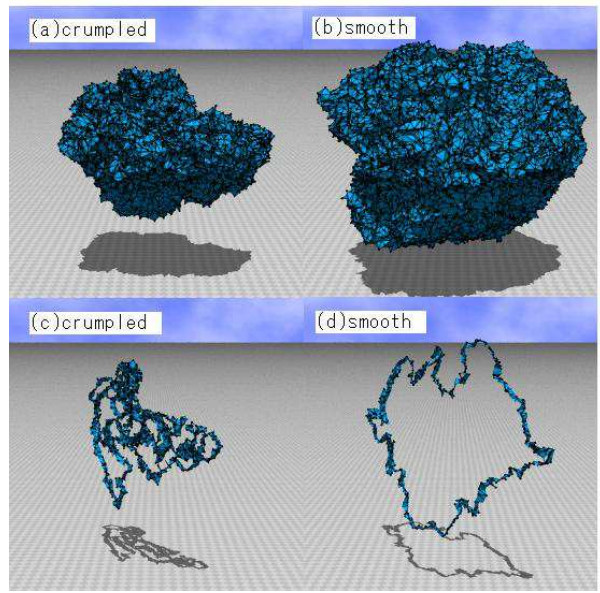}
 \caption{Snapshots of surfaces at (a) the crumpled phase and at (b) the smooth phase,  and  (c) the section of the surface in (a), and (d) the section of the surface in (b). The snapshots were obtained at $b\!=\!0.769$ on the surface of $N\!=\!15212$. The mean square size is about (a) $X^2\!=\!54$ and (b) $X^2\!=\!138$.}
\label{fig-7}
\end{figure}
Figure \ref{fig-7}(a) is a snapshot of the $N\!=\!15212$ surface in the crumpled phase at $b\!=\!0.769$, and Fig. \ref{fig-7}(b) is the one in the smooth phase at the same $b$. The mean square size is about $X^2\!=\!54$ and $X^2\!=\!138$ in Figs. \ref{fig-7}(a) and \ref{fig-7}(b), respectively. The sections for the surfaces in Figs. \ref{fig-7}(a) and \ref{fig-7}(b) are depicted in Figs. \ref{fig-7}(c) and \ref{fig-7}(d) respectively. Surfaces are rough in short scales even in the smooth phase shown in Figs. \ref{fig-7}(b), whereas they are smooth in the long range scale. Surfaces rough in short scales can also be seen deep in the smooth phase. There appears to be only a spherical mono-layer surface in the smooth phase, and there are no apparent oblong, linear or branched polymer surfaces for $N\leq 15212$ at least. On the contrary, surfaces in the crumpled phase at the transition point are not completely collapsed, and they appear not only crumpled but also smooth. Thus, we understand a reason why the Hausdorff dimension $H_{\rm cru}$ in Eq. (\ref{H}) is less than the physical bound.  

\begin{figure}[hbt]
\centering
\includegraphics[width=8.5cm]{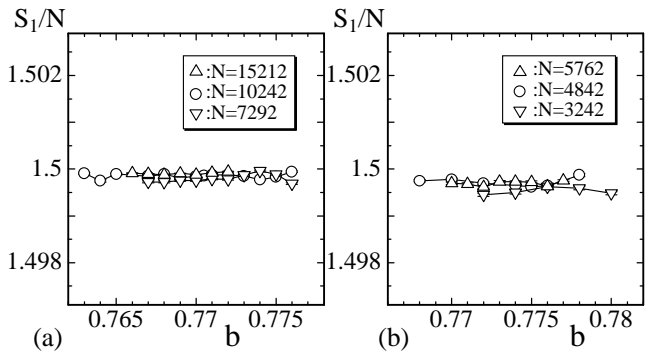}
 \caption{$S_1/N$ against $b$ obtained on surfaces of (a) $N\!=\!15212$,  $N\!=\!10242$,  $N\!=\!7292$, and (b) $N\!=\!5762$,  $N\!=\!4842$,  $N\!=\!3252$. }
\label{fig-8}
\end{figure}
Finally, we plot $S_1/N$ against $b$ in Figs. \ref{fig-8}(a) and \ref{fig-8}(b) in order to check that the equilibrium configurations were obtained in the MC simulations. We see that the expected relation  $S_1/N\!=\!3(N\!-\!1)/(2N) \!\simeq\!3/2$ holds in all cases in the figure. The deviations are very small. It was also confirmed that the relation $S_1/N\!\simeq\!3/2$ is satisfied in all other cases, which were not presented in the figure.

\section{Summary and conclusions}
To conclude, we have shown that the tethered surface model of Helfrich and Polyakov-Kleinert undergoes a first-order phase transition between the smooth and the crumpled phases by performing canonical MC simulations on spherical surfaces of size up to $N\!=\!15212$. The surfaces were constructed by dividing the icosahedron. The discrete form of the Hamiltonian, a linear combination of the Gaussian term and the bending energy term, is the one widely used in the numerical studies carried out so far. The first-order transition was observed on surfaces of size $N\!\geq\!7292$. We have checked that the transition appears as a second-order one on relatively small surfaces, and hence that our results are consistent with results reported in previous numerical studies. We also confirmed that the Hausdorff dimension $H_{\rm smo}$ is close to the topological dimension $H\!=\!2$ in the smooth phase at the transition point, and that $H_{\rm cru}$ remains finite and is less than the physical bound $H\!=\!3$ in the crumpled phase at the transition point. Consequently, it is possible to consider that the model represents a smooth and crumpled state in real physical membranes at the crumpled phase close to the transition point. Further numerical studies would clarify this problem. Since the first-order nature of the transition has been confirmed in the HPK model, it would be interesting to study the model in more detail with large scale simulations: the phase diagram in the tensionless model, the dependence of the transition on the topology of surfaces, and the phase diagram of the fluid surfaces.  

The simulations were done by using a Pentium-4 (2.8GHz--3.2GHz) PC with an Intel Fortran Compiler for Linux and one for Windows. The total number of CPU time was about 1350 days. Snapshots of surfaces were generated with POV-Ray for Windows v3.5.

\acknowledgements
This work is supported in part by a Grant-in-Aid for Scientific Research Grant No. 15560160. 




\begin{thebibliography}{00}





\bibitem{HELFRICH-1973}
 W. Helfrich, Z. Naturforsch, 28c (1973) 693.

\bibitem{POLYAKOV-NPB1986}
 A.M. Polyakov, Nucl. Phys. B 268 (1986) 406;\\
H. Kleinert, Phys. Lett. B 174 (1986) 335.

\bibitem{NELSON-SMMS2004}
D. Nelson, in {Statistical Mechanics of Membranes and Surfaces, Second Edition}, edited by  D. Nelson, T.Piran, and S.Weinberg, (World Scientific, 2004), p.1. 

\bibitem{David-TDQGRS-1989}
F. David,  in {Two dimensional quantum gravity and random surfaces, Vol.8}, edited by  D. Nelson, T. Piran, and S. Weinberg, (World Scientific, Singapore, 1989), p.81.

\bibitem{Wiese-PTCP2000}
K. Wiese, in: C.Domb, J.Lebowitz  (Eds.), Phase Transitions and Critical Phenomena, Vol. 19, Academic Press, London, 2000, p.253.

\bibitem{Bowick-PREP2001}
 M. Bowick and A. Travesset, Phys. Rep. 344 (2001) 255.

\bibitem{WHEATER-JP1994}
 J.F. Wheater, J. Phys. A Math. Gen. 27, (1994) 3323.

\bibitem{Peliti-Leibler-PRL1985}
 L. Peliti and S. Leibler, Phys. Rev. Lett. 54 (15)  (1985) 1690.

\bibitem{David-EPL1986}
 F. David, Europhys. Lett. 52 (8) (1986) 577.

\bibitem{DavidGuitter-EPL1988}
 F. David and E. Guitter, Europhys. Lett,  5 (8)  (1988) 709.

\bibitem{BKS-PLA2000}
 M.E.S. Borelli, H. Kleinert, and Adriaan M.J. Schakel, Phys. Lett. A 267 (2000) 201.

\bibitem{BK-PRB2001}
 M.E.S. Borelli and H. Kleinert, Phys. Rev. B 63, (2001) 205414. 


\bibitem{KANTOR-NELSON-PRA1987}
 Y. Kantor and  D.R. Nelson, Phys. Rev. A 36  (1987) 4020.

\bibitem{KANTOR-SMMS2004}
Y. Kantor, in {Statistical Mechanics of Membranes and Surfaces, Second Edition}, edited by  D. Nelson, T.Piran, and S.Weinberg, (World Scientific, 2004), p.111. 

\bibitem{WHEATER-NPB1996}
J.F. Wheater, Nucl. Phys. B 458 (1996) 671

\bibitem{BCFTA-JP96-NPB9697}
 M. Bowick,  S. Catterall,  M. Falcioni,  G. Thorleifsson, and K. Anagnostopoulos, 
J. Phys. I France  6 (1996) 1321;\\
 M. Bowick,  S. Catterall,  M. Falcioni,  G. Thorleifsson, and K. Anagnostopoulos, 
Nucl. Phys. Proc. Suppl. 47 (1996) 838;\\
 M. Bowick,  S. Catterall,  M. Falcioni,  G. Thorleifsson, and  K. Anagnostopoulos, 
Nucl. Phys. Proc. Suppl. 53 (1997) 746.

\bibitem{KY-IJMPC2000-2}
 H. Koibuchi and  M. Yamada, Int. J. Mod. Phys. C 11 (8) (2000) 1509.

\bibitem{KOIB-PLA2003-2}
 H. Koibuchi, N. Kusano, A. Nidaira, K. Suzuki, and T. Suzuki,  Phys. Lett. A 314 (2003) 1.

\bibitem{PKN-PRL1988}
M. Paczuski, M. Kardar, and D. R. Nelson, Phys. Rev. Lett. {\bf 60}, (1988)  2638.

\bibitem{KD-PRE2002}
J-P. Kownacki and H. T. Diep, Phys. Rev. E {\bf 66},  (2002)  066105.

\bibitem{Koibuchi-PRE-2004-1}
H. Koibuchi, N. Kusano, A. Nidaira, K. Suzuki, and M. Yamada, Phys. Rev. E, {\bf 69}, 066139 (2004).
\bibitem{CATTERALL-NPBSUP1991}
 S.M. Catterall, J.B. Kogut, and R.L. Renken, 
Nucl. Phys. Proc. Suppl. B 99A, (1991) 1.

\bibitem{AMBJORN-NPB1993}
 J. Ambjorn, A. Irback, J. Jurkiewicz, and B. Petersson, 
Nucl. Phys. B 393, (1993) 571.

\bibitem{ABGFHHM-PLB1993}
 K. Anagnostopoulos, M. Bowick, P. Gottington, M. Falcioni, L. Han, G. Harris, and E. Marinari, 
Phys. Lett. B 317 (1993) 102.

\bibitem{BCHHM-NPB9393}
 M. Bowick, P. Coddington, L. Han, G. Harris, and E. Marinari, 
Nucl. Phys. Proc. Suppl. 30 (1993) 795;\\
 M. Bowick, P. Coddington, L. Han, G. Harris, and  E. Marinari, 
Nucl. Phys. B 394 (1993) 791.

\bibitem{KY-IJMPC2000-1}
 H. Koibuchi and M. Yamada, Int. J. Mod. Phys. C11 (3), (2000) 441.

\bibitem{KOIB-PLA200234}
 H. Koibuchi,  Phys. Lett. A 300 (2002) 582; \\
 H. Koibuchi, N. Kusano, A. Nidaira, K. Suzuki, and M.Yamada,  Phys. Lett. A 319 (2003) 44; \\
 H. Koibuchi, N. Kusano, A. Nidaira, and K. Suzuki,  Phys. Lett. A 332 (2004) 141.


\bibitem{David-Wiese-PRL96}
F.David and  K.J.Wiese, Phys. Rev. Lett. {\bf 76}, (1996) 4564.
\bibitem{Gross-PLB1984}
D.J. Gross, Phys. Lett. B 138 (1984) 185.

\bibitem{Duplantier-PLB1984}
B. Duplantier,  Phys. Lett. B 141 (1984) 239.

\bibitem{JK-PLB1984}
J. Jurkiewicz and A. Krzywicki, Phys. Lett. B 148 (1984) 148.

\bibitem{Matsumoto-Nishimura-1998}
M. Matsumoto and T. Nishimura, "Mersenne Twister: A 623-dimensionally equidistributed uniform  pseudorandom number generator", ACM Trans. on Modeling and Computer Simulation Vol. 8, No. 1,  January (1998) pp.3-30.


\end{thebibliography}
\end{document}